\documentstyle[epsbox]{tmu}
\begin{document}

\title{%
Quark Confinement and Deconfinement Transition in Large Flavor QCD}

\author{%
      Kei-Ichi Kondo                   \\
{\it  Department of Physics, Faculty of Science, Chiba University,
 Chiba 263-8522, Japan. 
kondo@cuphd.nd.chiba-u.ac.jp}\\
}

\maketitle

\section*{Abstract}

We give a new criterion of quark confinement/deconfinement by deriving a low-energy effective theory of QCD.  The effective theory can explain Abelian dominance in low-energy physics of QCD, especially, quark confinement.  Finally, we apply the above criterion to the large flavor QCD and discuss its phase structure.  The result suggests the existence of conformal phase.

\section{Introduction}

The purpose of this talk is to give a new criterion of quark confinement and to apply this criterion to large flavor QCD for discussing the phase structure.  
The new criterion of quark confinement can be derived by combining the Abelian-projected effective gauge theory of QCD derived by the author \cite{KondoI} and a novel reformulation of gauge theory (as a deformation of a topological field theory) proposed recently by the author \cite{KondoII,KondoIII,KondoIV,KondoV,KondoVI}.
For these purposes, the maximal Abelian (MA) gauge fixing is essential.  This is based on the idea of 't Hooft \cite{tHooft81}, Abelian projection.  Immediately after this proposal, the hypothesis of Abelian dominance in low-energy physics of QCD was proposed by Ezawa and Iwazaki \cite{EI82}.
In the MA gauge \cite{KLSW87}, actually, Abelian dominance was discovered by Suzuki and Yotsuyanagi \cite{SY90} a decade ago based on numerical simulation on a lattice and has been confirmed by the subsequent simulations by various authors, see \cite{review} for reviews.

\par
In the paper \cite{KondoI}, we have constructed an effective Abelian gauge theory which is considered to be valid in the low-energy region of QCD.  We called it the Abelian-projected effective gauge theory (APEGT), although this name is somewhat misleading as will be explained below.
Before this work, a number of low-energy effective gauge theories were already proposed based on the idea of Abelian-projection.  However, we should keep in mind that these models are constructed by ignoring all the off-diagonal gluon fields from the beginning by virtue of the Abelian dominance and/or by assuming the Abelian electro-magnetic duality, even if they can well describe some features of confinement physics in QCD phenomenologically.  In fact, they could not be derived starting with the QCD Lagrangian.  Therefore, one can neither answer how the off-diagonal gluon fields influence the low-energy physics nor why and how the electro-magnetic duality should appear from the non-Abelian gauge theory.
\par
In contrast to these models, the APEGT is the first principle derivation of effective theory from QCD.   It was shown that the off-diagonal gluons do affect the low-energy physics in the sense that off-diagonal gluons renormalize the effective Abelian gauge theory and let the coupling constant of effective Abelian gauge theory  run according to the renormalization group $\beta$-function which is exactly the same as the original QCD, thereby, exhibiting the asymptotic freedom. In this sense, the APEGT reproduces a characteristic feature of the original QCD, asymptotic freedom, even if it is an Abelian gauge theory.
Moreover, it was demonstrated how the dual Abelian gauge theory (magnetic theory) can in principle be obtained in the low-energy region of QCD.  In fact, monopole condensation leads to a dual Ginzburg-Landau theory supporting the dual superconductor picture of QCD vacuum.  
On the other hand, a version of the non-Abelian Stokes theorem shows that the Wilson loop operator can be expressed in terms of diagonal gluon fields, see e.g. \cite{KT99}.
Combining these results, we can explain the Abelian dominance in quark confinement.  
\par

\section{QCD in Maximal Abelian gauge}

We consider the Cartan decomposition of the gauge potential into the diagonal and off-diagonal parts,
\begin{equation}
  {\cal A}_\mu(x) = {\cal A}_\mu^A(x) T^A = a_\mu^i(x)T^i + A_\mu^a(x) T^a ,
\end{equation}
where $A=1,\cdots,N^2-1$ and $i=1,\cdots,N-1$ for the gauge group $G=SU(N)$.
Then we define the functional,
\begin{equation}
  R[A] := \int d^4x {1 \over 2} A_\mu^a(x) A_\mu^a(x) .
\end{equation}
The maximal Abelian (MA) gauge is obtained by minimizing the functional $R[A^U]$ w.r.t. the local gauge transformation $U(x)$ of $A_\mu^a(x)$.
Then we obtain the differential form of the MA gauge,
\begin{equation}
  \partial_\mu A_\mu^a - g f^{abi} a_\mu^i A_\mu^b := D_\mu^{ab}[a]A_\mu^b = 0 .
\end{equation}
This is nothing but the the background-field gauge with the background field $a_\mu^i$.  After the MA gauge is adopted, the original gauge group $G=SU(N)$ is broken to the maximal torus group $H=U(1)^{N-1}$.
The MA gauge is a partial gauge fixing which fixes the gauge degrees of freedom for the coset space $G/H$.
\par
The action of QCD in the MA gauge is given by 
\begin{eqnarray}
  S &:=& S_{YM} + S_{GF+FP} + S_{F} ,
\\
  S_{YM} &=& \int d^4x {-1 \over 4} F_{\mu\nu}^A F^{\mu\nu}{}^A ,
\\
  S_{GF+FP} &=& - \int d^4x  i \delta_B [\bar C^a (D_\mu[a]A^\mu+{\alpha \over 2}B)^a ] ,
\\
 S_{F} &=& \int d^4x \bar \Psi i \gamma^\mu (\partial_\mu - i g {\cal A}_\mu) \Psi ,
\end{eqnarray}
where $\delta_B$ is the BRST transformation and $\alpha$ is the gauge fixing parameter.  The gauge fixing for the residual symmetry $H$ will be discussed below.

\section{APEGT as a low-energy effective theory of QCD}

In order to obtain the effective theory which is written in terms of the diagonal fields $a^i, B^i, C^i, \bar C^i$ alone, we intend to integrate out all the off-diagonal fields $A^a, B^a, C^a, \bar C^a$.  We call such an effective field theory obtained in this way the abelian-projected effective gauge theory (APEGT).
That is to say, the APEGT is defined as
\begin{eqnarray}
  Z_{QCD} &:=& \int [d{\cal A}_\mu^A][dC^A][d\bar C^A][dB^A] \exp (i S_{QCD}) 
\nonumber\\
&=& \int [da_\mu^i][dC^i][d\bar C^i][dB^i] \exp (i S_{APEGT}) .
\end{eqnarray}
where
\begin{equation}
  \exp (i S_{APEGT}) 
= \int [dA_\mu^a][dC^a][d\bar C^a][dB^a] \exp (i S_{QCD}) .
\end{equation}
In the process of deriving the APEGT, we have introduced the anti-symmetric auxiliary tensor field $B_{\mu\nu}$ to avoid the difficulty caused by the quartic self-interactions among the off-diagonal gluons.  Here $B_{\mu\nu}$ is invariant under the residual gauge transformation $H=U(1)^{N-1}$.
First of all, we have integrated out $A_\mu^a$ and $B^a$ and obtained 
\begin{eqnarray}
  {\cal L} &=& -{1+z_a \over 4g^2} f_{\mu\nu}^i f^{\mu\nu}{}^i  
-{1+z_b \over 4} g^2 B_{\mu\nu}^i B^{\mu\nu}{}^i 
+ {z_c \over 2}  B_{\mu\nu}^i \tilde f^{\mu\nu}{}^i 
\nonumber\\
&+& i \bar C^a D_\mu[a]^{ac} D_\mu[a]^{cb} C^b
+ ({\rm ghost~self-interaction~terms})
\nonumber\\
&+& ({\rm higher-derivative~terms})
\end{eqnarray}
where $f_{\mu\nu}^i$ is the Abelian field strength 
$f_{\mu\nu}^i := \partial_\mu a_\nu^i - \partial_\nu a_\mu^i$ and $\tilde f_{\mu\nu}^i$ is the Hodge dual of $f_{\mu\nu}^i$.
This result shows that the off-diagonal gluons can not be ignored and that they influence the APEGT in the form of renormalization of the Abelian theory.  In fact, the renormalization factors $z_a, z_b, z_c$ are given by
$
 z_a = - {20 \over 3}N {g^2 \over (4\pi)^2} \ln {\mu \over \mu_0} ,
 z_b = + 2N {g^2 \over (4\pi)^2} \ln {\mu \over \mu_0} ,
 z_c = + 4N {g^2 \over (4\pi)^2} \ln {\mu \over \mu_0} .
$
In what follows we neglect the higher-derivative terms because we need the low-energy effective theory.  Moreover, we take into account a term $i \bar C^a D_\mu[a]^{ac} D_\mu[a]^{cb} C^b$ for a moment leaving the ghost self-interaction terms untouched.  This term leads to an additional renormalization for the Abelian field strength as,
\begin{eqnarray}
&&  \int [dC^a][dC^a] \exp \{ i \int d^4x i \bar C^a D_\mu[a]^{ac} D_\mu[a]^{cb} C^b \} 
\nonumber\\
&=& \exp [ \ln \det (D_\mu[a] D_\mu[a])] 
= \exp \{ i \int d^4x {z_d \over 4g^2} f_{\mu\nu}^i f^{\mu\nu}{}^i \} .
\end{eqnarray}
Thus we obtain the effective theory,
\begin{eqnarray}
  {\cal L} &=& -{1 \over 4g^2(\mu)} f_{\mu\nu}^i f^{\mu\nu}{}^i  
-{1+z_b \over 4} g^2 B_{\mu\nu}^i B^{\mu\nu}{}^i 
+ {z_c \over 2}  B_{\mu\nu}^i \tilde f^{\mu\nu}{}^i 
\nonumber\\
&+& ({\rm ghost~self-interaction~terms}) + ({\rm higher-derivative~terms})
\end{eqnarray}
where we have defined $g(\mu) := Z_a^{1/2} g$ with
$Z_a := 1-z_a+z_d= 1+ {22 \over 3}N {g^2 \over (4\pi)^2} \ln {\mu \over \mu_0} $.
A remarkable fact is that the gauge coupling constant $g(\mu)$ runs according to the same $\beta$-function as the original Yang-Mills theory,
\begin{equation}
  \beta(g) := \mu {dg(\mu) \over d\mu} = - b_0 g^3(\mu) ,
\quad b_0 = {11 \over 3}N > 0 .
\end{equation}
For $G=SU(2)$, this fact was first shown by Quandt and Reinhardt \cite{QR98} for $\alpha=0$ and subsequently by myself \cite{KondoI} for $\alpha \not= 0$.  The generalization to $SU(N)$ is straightforward.
So the APEGT is an effective Abelian gauge theory exhibiting the asymptotic freedom.
The $B_{\mu\nu}^i$ field is important to derive the dual Abelian gauge theory which leads to the dual superconductivity.  But we don't discuss this aspect, see \cite{KondoI}.  
The effect of dynamical quarks can be included into this scheme by integrating out the quark fields.  It results in further renormalization leading to the $\beta$-function with $b_0={11 \over 3}N - {4 \over 3}f r_F$ where $f$ is the number of quark flavors and $r_F$ is the dimension of fermion representation.
\par
In the MA gauge, it is believed that the off-diagonal gluons become massive, while the diagonal gluons behave  in a complicated way.  The massiveness of off-diagonal gluons has been confirmed by Monte Carlo simulations on a lattice \cite{AS99}.  An analytical explanation was given at least in the topological sector based on the dimensional reduction of the topological sector to the two-dimensional coset $G/H$ nonlinear sigma (NLS) model \cite{KondoII}.
In view of this fact, the integration of massive off-diagonal gluon fields can be interpreted as a step of the Wilsonian renormalization group.
In this sense, the APEGT obtained in this way is regarded as the low-energy effective theory describing the physics in the length scale $R>m_A^{-1}$ or in the low-energy region $p<m_A$.

\section{A novel reformulation of compact Abelian gauge theory}

Although the Lagrangian of APEGT is apparently written in the non-compact form, it should be regarded as a compact Abelian gauge theory based on the compact Abelian group $H$, since $H$ is embedded in the compact non-Abelian group $G$.
The compact $U(1)$ theory can have the topological nontrivial configuration as suggested from 
$\pi_1(U(1))={\bf Z} = 0, \pm1, \pm2, \cdots$ which corresponds to the winding number of the "path" around the circle $S^1$.
Unfortunately, we don't know any direct or explicit representation for the continuum version of compact or periodic Abelian gauge theory, while the lattice formulation is well known, e.g., Wilson, Villain and so on.
So, we try to incorporate at least the topological nontrivial contribution via the gauge fixing term based on \cite{KondoIII}.  We separate the degrees of freedom via the compact U(1) gauge transformation of the Abelian gauge potential as
\begin{equation}
 a_\mu(x) = v_\mu(x) + \omega_\mu(x), \quad 
\omega_\mu(x) := {i \over g} U(x) \partial_\mu U^\dagger (x) ,
\quad U(x) = e^{i\varphi(x)} \in U(1) ,
\end{equation} 
where we identify $v_\mu(x)$ with the topological trivial (perturbative) piece and $\omega_\mu(x)$ with the topological nontrivial (non-perturbative) piece. The partition function is rewritten into the form,
\begin{eqnarray}
  Z = \int [dU][dB][dC][d\bar C] \exp (iS_{TFT}) 
\int [dv_\mu][d\tilde B][d\tilde C][d\bar {\tilde C}] \exp(iS_{p}) , 
\end{eqnarray}
where
\begin{eqnarray}
  S_{p} &=& \int d^4x \{ -{1 \over 4}(\partial_\mu v_\nu - \partial_\nu v_\mu)^2
- i \tilde \delta_B[\bar {\tilde C}(\partial_\mu v^\mu + {\alpha \over 2}\tilde B)] \} ,
\\
 S_{TFT} &=& \int d^4x i \delta_B \bar \delta_B ({1 \over 2}\omega_\mu \omega^\mu
+ i C \bar C) .
\end{eqnarray}
Here $\delta_B (\bar \delta_B)$ is the BRST (anti-BRST) transformation for the fields $\omega_\mu, B, C, \bar C$ in the topological sector,
\begin{equation}
\delta_B \omega_\mu = \partial_\mu C, \quad 
\delta_B B = 0, \quad 
\delta_B C = 0, \quad 
\delta_B \bar C = iB ,
\end{equation}
\begin{equation}
\bar \delta_B \omega_\mu = \partial_\mu \bar C, \quad 
\bar \delta_B \bar B = 0, \quad 
\bar \delta_B C = i \bar B, \quad 
\bar \delta_B \bar C = 0
\end{equation}
where $B+\bar B = 0$.
On the other hand, $\tilde \delta_B$ is the BRST transformation for $v_\mu, \tilde C, \bar{\tilde C}, \tilde B$ in the perturbative non-topological sector.  
\begin{equation}
\tilde \delta_B v_\mu = \partial_\mu \tilde C, \quad 
\tilde \delta_B \tilde C = 0, \quad 
\tilde \delta_B \bar{\tilde C} = i \tilde B,  \quad 
\tilde \delta_B \tilde B = 0.
\end{equation}
For the topological trivial sector, we adopt the Lorentz gauge $\partial_\mu v^\mu=0$.  For the topological nontrivial sector, on the other hand, we adopt the special choice which is similar to the MA gauge in its form (but the meaning is quite different).  

By making use of the superspace formalism due to Bonora and Tonin \cite{BT81}, we can translate the topological sector into the $OSp(4|2)$ invariant form \cite{KondoII}.  In this sense, the topological sector has a hidden supersymmetry.  Hence the dimensional reduction {\it a la} Parisi and Sourlas \cite{PS79} takes place.  As a result, the topological sector reduces to the O(2) NLSM in two dimensions \cite{KondoIII}.  Then we have the equivalence,
\begin{equation}
  Z_{YM} \cong Z_{APEGT} = Z_{TFT} Z_{p} = Z_{O(2)NLS} Z_{p} ,
\end{equation}
where the O(2) NLS model is given by
\begin{equation}
  Z_{O(2)NLS} = \int [dU] \exp (- S_{O(2)}),
\quad  S_{O(2)} := {\pi \over g^2} \int d^2z \partial_\mu \varphi(z) \partial_\mu \varphi(z) , 
\end{equation}
Here $[dU]$ is a Haar measure and $\varphi(z)$ is the periodic angle variable for U(1).  Hence the theory is not a free scalar theory due to periodicity.  The equation of motion, $\nabla^2 \varphi(z)=0$ (mod $2\pi$) has a solution of the type,
$\varphi(z) = \sum_{i} Q_i Im \ln (z-z_i)$ which has isolated singularities with a vorticity $Q_i$ at $z=z_i$.  
Taking into account the vortex configuration, the O(2) NLSM is rewritten into the neutral ($\sum_{i} Q_i=0$) Coulomb gas,
\begin{equation}
  Z_{O(2)NLS} = \sum_{n=0}^{\infty} {\zeta^n \over (n!)^2} \int \prod_{j=1}^{n} d^2z_j \exp \left[ (2\pi)^2 \beta \sum_{i,j} Q_i Q_j {1 \over 2\pi} \ln {R_0 \over |z_i-z_j|} \right] ,
\end{equation}
where $\beta := 2\pi/g_{YM}^2$ and $\zeta=\exp(-S^{(1)})$ is the fugacity coming from the self-energy (action) of vortices $S^{(1)}$. It is known \cite{Coleman75} that the Coulomb gas is equivalent to the sine-Gordon model in two dimensions.  A pair of vortex and anti-vortex forms a dipole.  It is known that the weak coupling phase ($2\pi\beta>4$ ,i.e., $g_{YM}<\pi$) is an ordered phase of dipoles, whereas the strong coupling phase ($g_{YM}>\pi$) is the disordered phase where the dipole melts and free vortices form a plasma.   The Wilson loop integral $\oint_C dx^\mu \omega_\mu = {2\pi \over g} \sum_{i} Q_i$ is given the sum of the vorticity of each vortex which is inside the Wilson loop has. The ordered phase corresponds to the deconfining phase, since the sum of vorticities is zero.
In the disordered phase the Wilson loop can have a area law \cite{KondoIII}.
A phase transition separating two phases occurs, although there is no local order parameter for this transition in agreement with Coleman's theorem \cite{CMW}. The phase transition is called the Berezinskii-Kosterlitz-Thouless (BKT) transition \cite{BKT}.  Therefore, we conclude that the APEGT has a strong coupling confining phase for $\alpha(\mu):={g_{YM}^2(\mu) \over 4\pi} > {\pi \over 4}$.
The running coupling constant increases monotonically as the energy scale $\mu$ decreases and eventually reaches the critical coupling $\alpha_c$ at sufficiently low energy or sufficiently long distance.  Hence we propose a criterion of quark confinement, 
\begin{equation}
  \alpha(\mu) > \alpha_c = {\pi \over 4}  .
\end{equation}

\section{Large flavor QCD}

\begin{figure}[t]
\begin{center}
\epsfile{file=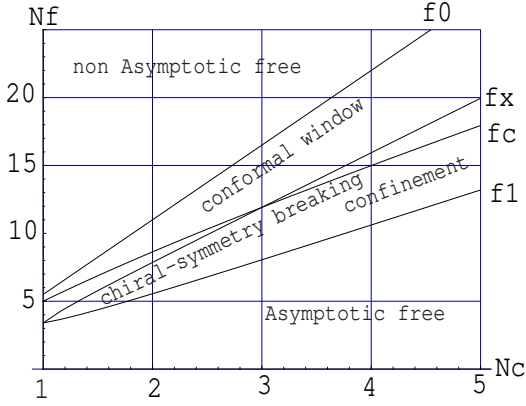,width=0.50\textwidth}
\end{center}
\caption{Conjectured phase diagram of large flavor QCD with $N_c$ colors and $N_f$ flavors.}
\label{fig.1}
\end{figure}

The RG $\beta$ function in SU(N) QCD with $f$ quark flavors is given by
\begin{eqnarray}
  \beta(\alpha) := \mu {\partial \alpha(\mu) \over \partial \mu} 
= - c_0 \alpha^2 - c_1 \alpha^3 + O(\alpha^4) ,
\\
c_0 = - {1 \over 3\pi}(f-f_0), \quad f_0 := {11 \over 2}N,
\\
c_1 = - {13N^2-3 \over 24\pi^2 N} (f-f_1), \quad f_1 := {34N^3 \over 13N^2-3} .
\end{eqnarray}
Fixing the number of colors $N$, we find three characteristic behaviors depending on the number of flavors $f$:  
\begin{enumerate}
\item[(a)] $f>f_0$ ($c_0<0, c_1<0$): non asymptotic free (IR free) and $\alpha(0)=0$;
\item[(b)] $f_0>f>f_1$ ($c_0>0, c_1<0$): A nontrivial IR fixed point exists at $\alpha=\alpha^*:=-c_0/c_1$.  Hence $\alpha(0)=\alpha^*$;
\item[(c)] $0 \le f<f_1$ ($c_0>0, c_1>0$): asymptotic free and $\alpha(0)=\infty$. 
\end{enumerate}
Note that $\alpha^*$ increases as $f$ decreases in case (b). So, the confinement is realized below a critical number of flavors $f_c$ such that $\alpha^* \ge \alpha_c$.  Therefore the critical $f$ is given by
\begin{equation}
 f_c(N) = {2(22\pi+17N\alpha_c)N^2 \over 8\pi N+(13N^2-3)\alpha_c} .
\end{equation}
For $N=3$, the critical coupling $ \alpha_c = {\pi \over 4}$ happens to coincide with the critical coupling of chiral symmetry breaking  obtained by the Schwinger-Dyson equation within the ladder approximation,
\begin{equation}
 \alpha_\chi={2N \over N^2-1}{\pi \over 3}.
\end{equation}
 The spontaneous breaking of chiral symmetry occurs when $\alpha(0)>\alpha_\chi$. 
Thus we have $f_c(3)=12$ for $N=3$.  The obtained phase structure is consistent with the numerical result on a lattice \cite{IKKSY98} at least qualitatively, although the lattice results suggests a little bit smaller value, $f_c(3)=6$.  
This critical value can be lowered at least to $f_c(3) \cong 10$ using the 't Hooft renormalization scheme in which the two loop result is exact.  
For large colors $N>3$, the naive estimate shows that the chiral symmetry breaking occurs already at somewhat larger flavor $f_\chi$ than $f_c$ at which the confinement takes place.  In such a case, we must modify the above argument on confinement, since we have used the $\beta$-function which is calculated for the massless quark.  Once the chiral symmetry is spontaneously broken, the quark acquires a non-zero mass and hence the quark does not contribute to the $\beta$-function below this mass scale. Below $f_\chi$, confinement will be realized.
See Fig.1.

\section{Conclusion and discussion}

We have derived a low-energy effective theory of QCD by integrating out off-diagonal gluon fields and quark fields in the MA gauge.   The resultant theory (APEGT) is the Abelian (compact) gauge theory with a running coupling constant governed by the same $\beta$-function as the original QCD.  The APEGT can explain the Abelian dominance in quark confinement.
By making use of a novel reformulation of gauge theory, we derived a new criterion of quark confinement in QCD.  
\par
Then we applied this criterion to large flavor QCD and given a schematic phase diagram of QCD with $N$ colors and $f$ quark flavors.  The resultant phase diagram is similar to that obtained by Appelquist, Terning and Wijewardhana
\cite{ATW96} by taking into account the chiral symmetry
breaking/restoration. In sharp contrast with their approach,
we have taken into account the quark confinement to derive the phase
diagram.  However, our treatment of chiral symmetry breaking is
still insufficient in the sense that we can not treat the chiral
symmetry breaking and quark confinement on equal footing.  Our
result suggests the existence of conformal phase when $f_0>f>f_c$
without losing the asymptotic freedom.  The existence of an essential
singularity in the conformal phase transition claimed in \cite{MY97}
is compatible with our result, since the BKT transition exhibits
the essential singularity, e.g, 
\begin{equation}
  m \sim \Lambda \exp [- c(\alpha^*/\alpha_c-1)^{-1/2} ] .
\end{equation}
Our analysis suggests that confinement without chiral symmetry breaking might be possible for small color $N=2$.
Anyway, we need to analyze the quark confinement and chiral symmetry breaking in the same framework.  I hope to report any progress in this direction in near future.

\newpage


\end{document}